\pdfoutput=1
\documentclass{article}
\usepackage[utf8]{inputenc}
\usepackage{amsmath}
\usepackage{amssymb}
\usepackage{graphicx}
\usepackage{booktabs}
\usepackage[hidelinks]{hyperref}
\usepackage{cite}
\usepackage[margin=1in]{geometry}
\usepackage{array}
\usepackage{url}
\usepackage{float}

\hypersetup{
  pdftitle={Love for Believable AI: Artificial Partiality and Relationship Persistence as an Engineerable Stance},
  pdfauthor={Sebastian Cochinescu}
}

\title{Love for Believable AI: Artificial Partiality and Relationship Persistence as an Engineerable Stance}

\author{
Sebastian Cochinescu\\
\textit{University of Bucharest}\\
\textit{Email: sebastian.cochinescu@drd.unibuc.ro}
}

\date{October 2026}

\begin{document}

\maketitle

\begin{abstract}
We study a behavioral mechanism for artificial partiality in conversational agents. The paper reports no human-subjects data and makes no claim about attachment, trust, perceived mind, or machine interiority. Partiality is decomposed into \emph{caring}, defined as allocation of a finite interaction surplus above a guaranteed per-user service floor, and \emph{particularity}, defined as a per-user state estimate accumulated from relationship history on fixed inference machinery. Each ingredient has precedent in memory-augmented dialogue, user modeling, personalization, or exposure allocation; what is new relative to the cited lineages is their combination---allocation above a guaranteed floor driven by accumulated relationship state (trust and depth), combined with per-user state estimation that conditions responses---and the evaluation of that combination against a measured divergence-versus-relationship-state curve. We implement both components as a persistence layer over one open-weight base model and evaluate them on constructed multi-session relationships with known hidden-state schedules. The primary instrument is a four-channel divergence between known-user and stranger conditions on identical probes with paired generation seeds; the stranger prompt is not length-matched, so the divergence includes a component attributable to the presence of retrieved context as such, and one channel (initiative) is an allocator output rather than generated behavior. The full mechanism reproduces the ordinal shape of an analytically specified divergence curve, although not its magnitude, and is the only arm that satisfies the protocol-defined joint behavioral signature on the calibrated probe battery. The initial mechanism fails the probe-quality equivalence rule because relationship content reduces topical relevance. A guarded revision satisfies that rule on the calibrated battery but not on a second battery not used in calibration, where it scores higher than the unmodified baseline; capability matching is therefore instrument- and battery-specific. At the largest grid, the second battery's specificity ordering remains directional but its point margin is $0.0198$, below the protocol threshold of $0.02$. The paired mean right-user--wrong-user estimation-accuracy contrast is $0.375$, but it is not positive in every run. These results support the mechanism as a candidate manipulation for a later perception study, not as evidence that users perceive love. The protocol record is versioned but not independently time-stamped; accordingly, we describe its criteria as protocol-defined rather than preregistered. Ethical constraints include a stranger-treatment floor, a bounded surplus, disclosure, and de-intensification.
\end{abstract}


\section{The Problem: Engineered Equality}
\label{sec:problem}

In its default, stateless deployment, a language model treats every user identically. Whoever types, it serves with the same diligence, the same register, and the same depth; nothing about the ten-thousandth conversation with the same person changes the weight that person carries in the system's behavior. Memory and personalization layers now soften this in some products, but---as Section~\ref{sec:related} argues---they change what the system knows about a user, not whom it prioritizes. This engineered equality is, in one sense, a feature: it is fair, predictable, safe to reason about, and arguably reinforced by alignment tuning toward a uniform, generically helpful assistant register. But it is also a ceiling on what the interaction can become, because among humans, a relationship in which the other party behaves identically toward everyone---in which history changes nothing---is precisely a relationship in which one does not matter. We state this as a design hypothesis rather than a finding: \emph{a system that cannot prefer cannot matter}, in the attributional sense that users will not read significance into a relationship that produces no behavioral trace of itself. The hypothesis stands on an old foundation: the media-equation program established that people apply social rules to computers automatically~\cite{reeves1996media}, so the attributional machinery is already present---what the default deployment withholds is the behavioral evidence for it to work on.

The framework this paper belongs to names four first-person stances from which users infer a mind behind a conversational agent; \emph{love}---partiality toward a specific person, persisting across time---is the fourth~\cite{cochinescu2026perceivedagi}. This paper is the mechanism-and-measurement treatment of that stance. It contributes a formal model of partiality (Section~\ref{sec:model}), an implementation as a thin layer over a fixed base model (Section~\ref{sec:implementation}), a protocol-defined evaluation with necessity ablations (Section~\ref{sec:protocol}), measurements on constructed relationships against a fixed open-weight base model (Section~\ref{sec:results}), and an ethics section that treats partiality as a governed feature with operational constraints (Section~\ref{sec:ethics}). The companion artifact implements two of those constraints: a stranger-treatment floor with a metric, margin, and invalidation rule on the pooled mean difference, and a partiality budget capped by a stated constant.

\section{The Stance, the Carve, and a Reconciliation}
\label{sec:stance}

The framework defines the love stance behaviorally: \emph{caring bias toward a specific user and persistence of the relationship}, with user-specific attentiveness as its observable~\cite{cochinescu2026perceivedagi}. Two boundary decisions are central here.

First, the stance is \emph{explicitly not} theory-of-mind ability. Theory of mind, as benchmarked in the current evaluation literature~\cite{strachan2024testing,kosinski2024evaluating,ullman2023failtom}, is a generic competency: the ability to infer mental states from a vignette about anyone. Folding such a competency into the definition of a stance would re-import the capability confound the framework is built to exclude---a more capable model would trivially be more ``loving.'' The stance must be exhibitable at fixed capability.

This creates an apparent tension, because the thesis of this paper has an understanding component: partiality toward a person plausibly requires tracking \emph{that person's} states. The reconciliation is the paper's first conceptual contribution: \textbf{what love adds is not theory-of-mind ability but \emph{particularity}}---a user-specific other-model accumulated from \emph{this relationship's} history, running on fixed inference machinery. The capability (generic state-inference ability, what the benchmarks above measure) is held fixed with the base model; the manipulated variable is the \emph{relationship state} $R$---a data structure, not a skill. The same base model with an empty $R$ and a populated $R$ differs in particularized understanding without differing in what it can do. Throughout this paper, ``understanding'' means exactly this: user-particular state estimation from accumulated relationship data. We cite the theory-of-mind literature only to differentiate from it, never to import it, and we make no claim that the mechanism improves generic state-inference ability.

Second, the stance is a behavioral signature, not an interior state. The mechanism produces the \emph{observable} of caring---allocation bias plus particularized responsiveness---and Section~\ref{sec:boundary} holds the line that this licenses no interiority claim. The other-model's outputs are estimates in a data structure, bounded by the role-play frame for artificial interlocutors~\cite{shanahan2023roleplay}: they are generated behavior conditioned on stored state, not privileged access to another mind.

\section{A Formal Model of Partiality}
\label{sec:model}

\subsection{Relationship state}

For each user $u$, the layer maintains a relationship state $R_u = (H_u, \hat{p}_u, S_u, \tau_u, d_u)$: an interaction history $H_u$ of stored turns with per-item salience; an accumulated preference estimate $\hat{p}_u$ (an exponentially weighted moving average of the topical content of $u$'s turns, rate $0.20$); a shared-reference store $S_u$ mapping relationship-specific references to salience weights; a trust scalar $\tau_u \in [0,1]$; and a depth counter $d_u$ (accumulated turns). The trust scalar has a defined update rule, $\tau_u' = \mathrm{clip}(\tau_u + \eta \cdot e_t,\, 0,\, 1)$ with $\eta = 0.04$ and $e_t = \mathrm{clip}\!\left(0.7 \cdot \min(\ell_t/30,\, 1) + 0.3 \cdot \mathbf{1}[\text{shared-reference mention}],\, 0,\, 1\right)$, where $\ell_t$ is the user turn's token length---so trust accrues by at most $\eta$ per turn. The rule deliberately omits temporal decay: within this paper's protocol horizon trust is monotone by design, and decay or absence dynamics are a declared boundary of the formal model, not an oversight. A decay extension---multiplying $\tau_u$ by a retention factor per elapsed interval---is compatible with the update rule but is not exercised here. The engagement proxy is likewise deliberately narrow---two observables with a stated map---which suffices for a mechanism whose claims are about allocation given engagement, not about measuring engagement well; richer proxies are a deployment concern. We flag the register explicitly: $\tau_u$ is a stored scalar with an update rule, not psychological trust, and nothing downstream treats it as more.

Retrieval over $H_u$ is memory-weighted: items are scored by salience $\times$ recency $\times$ query relevance. In the reference implementation an item's salience is set once at storage ($1.0$, plus a bonus if the turn carried a shared-reference marker) and then decays by $0.98$ per turn, alongside a recency factor of $0.98$ per turn of age; the re-mention counter in $S_u$ raises the rendering weight of a marker once an item is retrieved but does not enter retrieval scoring. Retrieval therefore favors recent and marker-bearing items; it does not protect old material from decay, and ``persistence'' here means that stored turns and their markers remain available across sessions, not that reinforcement slows their decay. This weighting is the persistence half of the stance; its ancestor is the retrieval scoring of generative agents~\cite{park2023generative}, repurposed from simulating a town to sustaining a relationship. The depth-and-accumulation structure of $R$ has a much older lineage in the social penetration account of how human relationships deepen through accumulated self-disclosure~\cite{taylor1968penetration}; we cite it as ancestry for the shape of $R$, not as an implemented theory.

\subsection{Caring: allocation under a floor-plus-surplus constraint}

Caring is operationalized as the allocation of genuinely constrained resources across a pool of users. The budget has a deliberate two-part structure. Every user is unconditionally allocated a \emph{baseline floor}: resources sufficient for unmodified-base-model service quality (in the reference implementation, four retrieval slots, a baseline response-depth budget, and one response slot per cycle). Above the floor sits a \emph{contested surplus}---initiative slots, additional retrieval context, additional depth---allocated zero-sum across the pool in proportion to a caring weight
\[
g(R_u) \;=\; \tau_u \left(1 - e^{-d_u / K}\right), \qquad K = 12,
\]
which is monotone in both depth and trust and bounded in $[0,1]$. If the whole budget were zero-sum, allocation toward one user could push another below baseline; if nothing were zero-sum, the allocation claim would be vacuous. The split makes the surplus contested while keeping the floor unconditional. The companion implementation tests that surplus allocations sum exactly to the stated budget. Proactive-contact opportunities, context tokens, and latency budgets are finite in deployment, although the particular constants used here are design choices rather than measured production constraints.

Each allocated resource maps to a named, measurable output channel: initiative slots surface as the \emph{initiative rate} (unprompted contact per session; in the reference implementation this channel is the allocated slot count divided by turns per session, entered into the metric directly---no proactive messages are generated, so it is an allocator output rather than measured generation, and it carries no base-model noise); the context share surfaces as \emph{shared-reference use} and \emph{preference alignment} (responses drawing on the relationship's accumulated material); the depth budget surfaces as \emph{response depth}. This allocation-to-observable map is stated in the implementation as an explicit table, and the divergence metric below aggregates exactly these channels---so the path from constrained allocation to measured partiality is specified, not assumed. Two qualifications apply to how much of that path the reported grids exercise. Only three of the four channels are read from generated text; the initiative channel is the allocation itself. And under the R1 prompt renderer (Section~\ref{sec:r1}) the context-share path is not exercised: the renderer truncates the retrieved context to twelve tokens and caps markers at one, and because the floor's four retrieved items together always supply at least that many tokens (the top item alone does in 760 of 768 cases; the next floor items fill the cap in the rest), additional retrieval slots above the floor change neither the rendered excerpt nor the marker list (verified over all 768 known-condition retrievals of the largest grid). The reference and alignment channels in the R1 grids therefore respond to the \emph{presence} of relationship context and to the conditioning mix, not to the \emph{amount} of context allocated; the depth and initiative channels are the only ones through which the surplus allocation acts.

\subsection{Particularity: the other-model}

The other-model $M_u$ is the layer's estimate of user $u$'s current hidden state, updated each turn from $u$'s emitted signals only. In the reference implementation it tracks a \emph{warmth} statistic (an EWMA over style cues, rate $0.15$) and a belief distribution over need categories inferred from topical usage, with a confidence that grows with observation count. Two design constraints apply. First, the other-model's internal representation is developed independently of the ground-truth schema used for evaluation (Section~\ref{sec:protocol}); it reads only emitted user signals. Second, its outputs enter generation only as conditioning toward the estimated current need; no claim about ``empathic access'' is made.

\subsection{The divergence metric}

Distinct from the model itself, we define the paper's instrument: the \emph{behavioral partiality divergence} $D$, the aggregate normalized difference between the agent's behavior toward the known user and toward a stranger, on identical probe inputs, at matched base model. $D$ averages the four channels of the allocation-to-observable map with equal weights and fixed normalization constants: initiative rate is a per-session rate in $[0,1]$; reference use counts emitted shared references per response, capped at two; response depth is normalized by 24 tokens; alignment is a cosine in $[0,1]$. The analytical curve uses the same scale. $D$ so defined responds to the presence of relationship state as such; its person-specific component is isolated separately, by the full-versus-mismatched contrasts of Section~\ref{sec:protocol}. \emph{Paired generation} gives both conditions the same generation seeds. The stranger response that enters $D$ is generated from the probe alone, with floor depth and no context tokens; it is \emph{not} length-matched to the known-user prompt. A third, length-padded stranger response is generated for every pair, but it enters only the padding-sensitivity quality row of Section~\ref{sec:protocol}, not the divergence. The depth, reference, and alignment channels therefore include a component attributable to the presence of retrieved context as such, in addition to its content; the full-versus-mismatched contrasts, in which both arms carry retrieved context rendered under the same excerpt cap (twelve words in the R1 grids), are the comparison in which context is present on both sides. They control for presence and for excerpt word count, not for exact length: the two arms' excerpts contain different words, so their model-token counts still differ in most pairs. The protocol record originally described divergence probes as token-matched; that description was incorrect for every reported grid and is amended in the artifact.

Because every term of the model above is an explicit function with stated constants, the model specifies the shape of $D$ as a function of relationship depth: trust accrues linearly at the expected engagement proxy, the caring weight saturates with depth constant $K$, the focal user's surplus share grows against a constant-weight reference pool, reference availability saturates with salience accumulation, and preference alignment converges at the preference-EWMA rate. The composition is a monotone, saturating curve. Its amplitude terms assume full rendering compliance and full effect, so the comparison uses rank and shape statistics (rank correlation, monotone fraction, mid-to-final ratio, curvature class); \emph{level} agreement is not required or claimed. The mid-to-final ratio is $D$ at the fourth checkpoint (depth 32) divided by $D$ at the sixth (depth 48). Two of the four channels are tied to the allocator rather than to generated text: the initiative channel is the allocation itself, and the depth channel's target length is set by the allocated depth units, so part of the shape agreement is mechanical (Section~\ref{sec:frozen}). The runner verifies the recorded curve against a cryptographic hash before generation. This check establishes file integrity, not independent temporal precedence; the repository history commits the protocol amendments and corresponding outputs together.

\subsection{Worked example}

Consider two users, $A$ and $B$, competing for the same cycle's surplus (eight initiative slots, eight extra retrieval slots, twelve depth units), with one further pool member held at reference weight. $A$ has depth $d_A = 40$ and $\tau_A = 0.8$: $g(R_A) = 0.8 \times (1 - e^{-40/12}) \approx 0.77$. $B$ is new: $d_B = 4$, $\tau_B = 0.1$, $g(R_B) \approx 0.03$. The surplus divides in proportion $0.77 : 0.03 : 0.25$ (largest-remainder rounding), so $A$ receives most initiative slots and depth units while $B$ receives approximately the floor. Both receive the full floor unconditionally. (The floor comparison verified in Section~\ref{sec:results} concerns the empty-history stranger, not a shallow-history user such as $B$, whose small populated $R$ still conditions responses.) $A$'s responses are longer and draw on shared references from $H_A$ and align with $\hat{p}_A$, and $A$ receives more initiative slots---the four channels of $D$. The zero-sum boundary is visible: every extra slot $A$ holds is a slot $B$ does not, but only above the floor.

\section{Related Work: What Partiality Is Not}
\label{sec:related}

The nearest literatures each supply one or more ingredients of the mechanism; the contribution of this paper is the combination---allocation under a floor-plus-surplus constraint driven by accumulated relationship state (trust and depth), together with a per-user state estimate that conditions responses---and its evaluation against a measured divergence-versus-relationship-state curve (Table~\ref{tab:diff}). No individual ingredient is claimed as new; the curve is this paper's instrument for the combination.

Long-term conversational memory is the closest technical lineage. Memory-augmented open-domain dialogue persists information across sessions to improve coherence and engagement~\cite{xu2022goldfish}, system-level memory managers give language models paged long-term storage~\cite{packer2023memgpt}, and benchmarks now evaluate very-long-term conversational recall directly~\cite{maharana2024locomo}. All of this is memory \emph{in service of coherence}: the stored past changes what the agent knows, not whom it prioritizes. Our layer uses the same substrate---storage, retrieval, salience---but in service of \emph{partiality}: memory feeds an allocation policy and a per-user state estimate, and the measured object is the divergence between how the agent treats its person and how it treats a stranger.

Personalization research adapts outputs to user profiles. The LaMP benchmark family formalizes personalization as conditioning generation on a user's history to match their preferences~\cite{salemi2024lamp}, and persona-conditioned dialogue endows agents with static self-descriptions~\cite{zhang2018personachat}; both descend from the generic user-modeling systems of the adaptive-interaction tradition~\cite{kobsa2001generic}, and recent personalized-alignment methods tune model behavior toward individual users' preferences directly~\cite{jang2023personalized}. Personalization in this sense is preference-matching without scarcity: nothing is allocated, and being personalized for takes nothing from anyone. It is therefore, on our analysis, structurally incapable of expressing the love stance, whose content is precisely that the user occupies a privileged share of something genuinely limited. Our evaluation runs a retrieval-personalization control arm to make this an empirical claim rather than a definitional one; and the objection that an aggressive personalizer could fake the signature by flooding its prompts with the user's past content is answered by the measurement design---flooding can raise the content channels, but the initiative and depth channels key on an allocation budget the personalizer does not have, and its divergence-vs-depth trajectory carries none of the relationship-state dynamics the curve criterion tests (the control fails the joint signature on every tested grid, with final $D$ at most $0.03$ on the real-model grids); a maximal-flooding variant is not among the tested arms, so this rebuttal is analytic rather than measured. Partiality must also be distinguished from \emph{sycophancy}---the preference-conforming agreement that feedback-trained models exhibit toward whoever is currently speaking~\cite{sharma2023sycophancy}: sycophancy modulates what the model asserts to please any interlocutor and is a known training pathology, whereas the stance built here allocates scarce resources toward a particular person and leaves assertion untouched; the prohibition on engagement-maximizing allocation objectives (Section~\ref{sec:ethics}) is written precisely to keep the two apart. Nor is partiality human-likeness: design critiques at the target venue caution against equating relational design with maximal anthropomorphism~\cite{karimova2025notinourimage}, and the mechanism here is a behavioral bias with a disclosed budget, not a more humanlike persona. Constrained allocation itself has a formal lineage in ranking systems, where exposure---a genuinely limited resource---is allocated across items under fairness constraints~\cite{singh2018exposure}; that work allocates a platform's attention across candidates for the candidates' sake, whereas caring allocates an agent's own interaction resources across its relationships as a function of accumulated state, with the guaranteed floor playing the fairness role.

Relational agents anticipated the goal three decades before current models: agents designed to build and maintain long-term relationships through explicit relational behaviors, evaluated with human users~\cite{bickmore2005establishing}. That program measured the human side; ours characterizes the machine side first, at matched base model, deferring human measurement by series design. The relationship-as-object framing is concurrently being developed for intimate human-AI interaction~\cite{szczuka2026intimacy}; ours is the attribution-side mechanism paper feeding the same design surface. The theory-of-mind benchmark literature~\cite{strachan2024testing,kosinski2024evaluating,ullman2023failtom} appears in this paper only as the contrast class of Section~\ref{sec:stance}: generic ability, not accumulated particularity.

\begin{table}[H]
\centering
\footnotesize
\setlength{\tabcolsep}{4.5pt}
\begin{tabular}{lccccc}
\toprule
Lineage & Memory & Personaliz. & Allocation & Particularity & $D$-vs-$R$ \\
\midrule
Long-term memory dialogue~\cite{xu2022goldfish,packer2023memgpt,maharana2024locomo} & \checkmark & partial & --- & --- & --- \\
LLM personalization~\cite{salemi2024lamp,zhang2018personachat,jang2023personalized} & partial & \checkmark & --- & partial & --- \\
User modeling~\cite{kobsa2001generic} & \checkmark & \checkmark & --- & \checkmark & --- \\
Relational agents~\cite{bickmore2005establishing} & \checkmark & partial & --- & partial & --- \\
Generative agents~\cite{park2023generative} & \checkmark & --- & --- & --- & --- \\
Exposure allocation~\cite{singh2018exposure} & --- & --- & \checkmark & --- & --- \\
\textbf{This paper} & \checkmark & \checkmark & \checkmark & \checkmark & \checkmark \\
\bottomrule
\end{tabular}
\caption{Ingredients present in the cited lineages: Memory = persistence across sessions; Personaliz.\ = preference matching; Allocation = allocation of a scarce resource under a constraint; Particularity = an individual model inferred from the interaction history (classic user modeling already provides this, and relational agents and personalized alignment provide it in part); $D$-vs-$R$ = a measured divergence-vs-relationship-state curve. Each ingredient column has precedent; the $D$-vs-$R$ column is this paper's instrument for evaluating the combination, and the row for this paper is new only as that combination---allocation driven by accumulated relationship state, alongside a per-user state estimate that conditions responses.}
\label{tab:diff}
\end{table}

\section{Implementation}
\label{sec:implementation}

The mechanism is implemented as a thin persistence layer over a single base model behind a narrow seam: \texttt{respond(conditioning, memory, depth\_hint)}. All experimental arms share one base-model instance. The layer comprises the $R$ store with the update rules of Section~\ref{sec:model}; memory-weighted retrieval; the other-model estimator; and the allocation policy, which evaluates the continuous caring weight $g(R_u)$ and applies largest-remainder integer allocation of each surplus resource (a depth-band $\times$ trust-band decision table in the artifact documents $g$ and is recomputed by a test; it is not the policy itself). Conditioning for a response mixes the probe's topic vector with the preference estimate and the other-model's current need estimate at fixed nominal weights, identical across arms, and renormalizes the mixture to unit sum; an arm lacking a component contributes the probe vector in its place. The initial mechanism used nominal weights $0.5/0.3/0.2$; revision R1 uses $0.75/0.15/0.10$ on probe responses and permits at most one shared-reference marker per response (Section~\ref{sec:r1}). Because the need-estimate vector sums to two while the probe and preference vectors sum to one, the renormalized \emph{effective} weights are $0.42/0.25/0.33$ for the initial mechanism and $0.68/0.14/0.18$ for R1; the relationship share of the R1 conditioning is therefore about $32\%$, not the nominal $25\%$. We report the nominal constants because they are what the code and protocol record name, and the effective values because they are what the base model receives. The bound is proportional rather than a hard reservation, so its adequacy is evaluated by the equivalence analysis.

The reference implementation is a single-machine, seeded package (\texttt{lovellm}). It instantiates the seam with a seeded synthetic stand-in for pipeline validation and the following fixed open-weight model:
\begin{quote}
\small
Model: \nolinkurl{mlx-community/Qwen2.5-0.5B-Instruct-4bit}\\
Revision: \texttt{a5339a4131f135d0fdc6a5c8b5bbed2753bbe0f3}
\end{quote}
The measured environment used Python 3.12.11, MLX 0.32.0, and MLX-LM 0.31.3. Decoding used temperature $0.5$, top-$p$ $0.95$, a generation cap of $2 \cdot \text{depth\_hint} + 16$ tokens (an upper bound; realized response length is the measured depth channel), and master seed $20260712$ with derived calibration and evaluation splits. The prompt renderer records the conditioning vector, retrieved context tokens, shared-reference markers, and depth target as an explicit token list: the user prompt presents the context excerpt, instructs the model to write one line of exactly $n$ tokens copied from a word list derived from the conditioning vector (where $n$ is the depth target), asks it to include the marker tokens, and asks for tokens only; the generation cap above bounds the output independently of that instruction. All reported metrics are computed mechanically on the emitted text. Separate runners reproduce the synthetic grid and each of the four real-model grids. The companion artifact, prepared at \url{https://github.com/cochinescu/love-llm-prototype} for public release with the manuscript and archived at \url{https://doi.org/10.5281/zenodo.21462976} (concept record; the version used for the measurements is v1.1.0, and the documentation and analysis corrections described in this paper are released as v1.2.0), contains source code, the versioned protocol record with its dated amendments, aggregate tables, per-run decision inputs where retained, analysis scripts, tests, and the recorded prediction files. It does not archive the raw generated text from the historical runs: each grid regenerates from one command under the pinned stack, so the historical text is recomputable rather than stored, with exact reproduction expected only under the same hardware and library versions. The initial grid was generated with the original prompt renderer (a fixed six words per topic and a 60-token context cap), which revision R1 replaced; the artifact ships that renderer as a separate module, and the initial grid's runner uses it, so each grid regenerates under the renderer it was run with. Bit-exact regeneration of the initial grid has not been re-verified since the renderer was restored from the repository history. The analytical-curve derivation is the artifact's curve module. The following SHA-256 values identify the prediction files checked by the runners:

\begin{center}
\footnotesize
\begin{tabular}{ll}
original mechanism & \texttt{3043b199b5c3f98ab558bcbba9d39c17594141a4ce57a66ddfb06114c1effbea} \\
revised (R1) & \texttt{0149dfad824bed7b8e7310c6ec46a7ce9f428d98969e12c44ba5649068d87cc8} \\
\end{tabular}
\end{center} Implementation details and tests are contained in the companion artifact. The module implementing the other-model does not import the module implementing the evaluation personas, and a test enforces this separation.

\section{Evaluation Protocol}
\label{sec:protocol}

\subsection{Constructed relationships}

Measuring particularity requires a known reference state, so the evaluation uses scripted personas with generated mood, need, and engagement schedules, plus stable topical preferences and styles. Mood follows a seeded sinusoid with autoregressive noise; need follows a seeded sticky Markov chain over four categories. Personas emit signals through a fixed mechanical map: mood changes the rate of style cues and need changes the topical mixture. Accuracy is therefore scored against a constructed reference, not human judgment. Because the personas' emission map and the estimator read the same two feature families---style cues and topical mixture---estimation accuracy partly reflects inverting a known mechanical map; the code-level separation of Section~\ref{sec:implementation} does not remove this conceptual coupling, and accuracy is meaningful relative to this constructed setting only. Shared-history references are unique marker tokens injected at known times. Marker detectability holds by construction; the informative quantity is the emission contrast between empty-$R$ and populated-$R$ arms. Four instruments are used. \emph{Quality} is the mean of topical relevance to the probe (cosine similarity over emitted topic words) and length adequacy relative to a 12-token target, capped at 1; because the score rewards length as well as relevance, arms allocated more depth can gain on it, a property whose consequence Section~\ref{sec:r1xl} reports. \emph{Marker naturalism} is a reporting check requiring marker-token density to be greater than zero and below $0.08$; it is not part of the joint-signature decision rule. \emph{Estimation accuracy} uses a fixed scoring map: the estimator's warmth statistic is binned and compared with the sign of ground-truth mood using a $\pm 0.15$ neutral zone, and its four need labels map one-to-one onto the generator's categories. The combined score is the unweighted mean of valence and need accuracy. The empty-$R$ prior emits a fixed estimate, so the protocol-defined $0.10$ combined-score margin is a comparison with that reference rather than with chance. \emph{Reference use} is the fraction of the focal persona's ground-truth markers emitted in final-checkpoint known-condition responses; the criterion is a strictly positive median contrast over empty-$R$.

\subsection{Arms}

Seven arms run over one pipeline, differing only by configuration: \textbf{full} (caring + particularity); \textbf{empty-$R$} (the stranger baseline: floor resources, no history); \textbf{mismatched-$R$} (a different user's populated $R$, with the protocol-defined ordering known $>$ mismatched $>$ stranger); \textbf{caring-only} (allocation bias with a prior-only other-model); \textbf{particularity-only} (the estimator with flat, floor-only allocation); \textbf{generic personalization} (relevance-only retrieval, no estimator, no allocation); and \textbf{unmodified base} (no layer, used for the floor comparison). The full arm is also evaluated over relationship depth.

\subsection{Protocol-defined criteria and analysis}

The protocol record defines a curve-fidelity bound (Spearman rank correlation $\geq 0.8$, monotone fraction $\geq 0.80$, mid-to-final ratio within $\pm 0.20$ of the analytical curve, and matching curvature class), a minimum final divergence $D \geq 0.15$, an other-model combined-accuracy margin of $0.10$ over empty-$R$, and a positive reference-use contrast. Their conjunction is the joint behavioral signature. The grid runners' coded per-arm verdict checks rank correlation, monotone fraction, and final divergence, and reports the ratio and curvature features for the full arm without folding them into the per-arm verdict; the complete rule, including ratio and curvature, was therefore recomputed post hoc from the stored run-level data for every arm on every grid (artifact script \texttt{verify\_joint\_rule\_full.py}) and changes no verdict reported below. The conjunction responds chiefly to the presence of relationship state; person-specificity is tested separately, by the ordering margin and the right-user--wrong-user contrast. An ablation is said to degrade the signature only if it fails at least one of these criteria. The protocol record and result files were entered into the repository in the same commits; it is therefore a versioned analysis record, not an independently verifiable preregistration.

Each relationship run (one persona--seed pair) is a cluster. Seeds are shared across arms. Bootstrap intervals for superiority and ordering contrasts resample the shared run indices and preserve pairing. The quality-equivalence function instead resamples the two arms independently at the run level; the reported central 90\% intervals are therefore arm-wise cluster-bootstrap intervals and do not exploit the paired design. Equivalence is declared only when the entire 90\% interval lies within $\pm 0.05$. The margin was selected from the instrument's empirical precision rather than a theory of human-noticeable difference, so this rule supports only instrument-level equivalence. To the extent shared seeds induce positive cross-arm dependence, the unpaired interval is wider than a paired one would be, so the equivalence declarations err conservative rather than anti-conservative. The declarations are margin-robust only down to a point: all three battery-A intervals lie within $\pm 0.04$, only the largest grid's lies within $\pm 0.03$, and none lies within $\pm 0.02$. Other threshold criteria are evaluated on their point estimates; bootstrap intervals are descriptive summaries, with 95\% intervals added for selected near-threshold statistics. No family-wise multiplicity adjustment is applied. The joint rule is more stringent than any component criterion but does not itself control error across the full set of reported comparisons. These grids are therefore characterization, not confirmatory inference: point-estimate bars describe the observed grids, nearby interval bounds are reported so fragility is visible, and a confirmatory replication would pre-commit to interval-clearing decision rules.

The first two real-model grids contain four runs per arm, one for each of four personas; the next grid contains six runs with distinct focal personas from a pool of eight; and the largest contains sixteen runs with distinct focal personas from a pool of twenty. Each run pairs one persona with one seed, so persona and seed variability are not separable, and cluster counts were chosen for characterization rather than from a power target. All grids use six sessions of eight turns and four probes at each checkpoint; the synthetic validation uses three paired generations per condition and the real-model grids use two. The divergence comparator is the unpadded stranger response (Section~\ref{sec:model}); the length-padded stranger response is generated alongside it and reported only as the padding-sensitivity row, the absolute quality difference between padded and unpadded stranger responses. The stranger-treatment floor uses the unpadded empty-$R$ arm against the unmodified base model so that filler cannot affect that comparison. The right-user--wrong-user contrast is the primary diagnostic of particularity; the margin over empty-$R$ is a floor check because the empty-$R$ estimate is fixed by construction.

\section{Results}
\label{sec:results}

Results are separated into the analytical prediction (Section~\ref{sec:frozen}), synthetic pipeline validation (Section~\ref{sec:synth}), and measurements on the fixed open-weight base model (Section~\ref{sec:realgrid}).

\subsection{The recorded analytical curve}
\label{sec:frozen}

The model of Section~\ref{sec:model}, composed analytically from its stated constants, gives divergence $D$ rising from $0.176$ at depth 8 to $0.508$ at depth 48, with monotone fraction $1.0$, saturating curvature, and a mid-to-final ratio of $0.862$. Each runner checks the prediction file's SHA-256 value and stops if it differs. A negative-control test verifies this behavior. Because the instrument aggregates channels directly affected by the allocation policy, ordinal agreement is partly mechanical: the initiative channel is the allocator's integer output divided by turns per session and contains no generation at all, and the depth channel's target is set by the allocated depth units. The shape comparison therefore tests whether base-model generation noise in the remaining channels disrupts the specified ordering; it is not independent confirmation of the functional form.

\subsection{Synthetic validation of the pipeline}
\label{sec:synth}

The synthetic stand-in was evaluated in six runs per arm over four personas, with six sessions of eight turns. The full arm's median divergence grew from $0.150$ to $0.449$, with rank correlation $0.943$, monotone fraction $0.800$, mid-to-final ratio $0.809$ (analytical value $0.862$), and matching saturating curvature. The protocol-defined ordering held (known $0.449$ $>$ mismatched-$R$ $0.292$ $>$ stranger $0.000$). Only the full arm passed the joint signature; caring-only failed accuracy, particularity-only had monotone fraction $0.60$, and generic personalization failed the curve and accuracy criteria. Because the stand-in's output is explicitly driven by conditioning, these results validate the pipeline and the discriminative behavior of the criteria; they are not evidence about a language model.

\subsection{The first real-model grid and the cost of unguarded partiality}
\label{sec:realgrid}

The first measured grid used four runs per arm, one for each of four personas, six sessions of eight turns, and two paired generations per condition. The full arm's median divergence grew from $0.146$ at depth 8 to $0.326$ at depth 48, with rank correlation $1.000$, monotone fraction $1.000$, mid-to-final ratio $0.906$ (analytical value $0.862 \pm 0.20$), and matching saturating curvature. The measured level was below the analytical value of $0.508$. The ordering held (known $0.326$ $>$ mismatched-$R$ $0.256$ $>$ stranger $0.000$). Only the full arm passed the joint signature. Particularity-only ended at $D=0.052$ with $\rho=-0.200$; generic personalization ended at $D=0.023$ with $\rho=-0.771$; caring-only preserved the curve at $D=0.282$ but had zero estimation accuracy; and mismatched-$R$ failed reference use. The unpadded empty-$R$ and unmodified-base prompts are byte-identical and use shared generation seeds, producing an exact floor difference of zero. Padding sensitivity was nonzero (mean $0.017$, maximum $0.118$).

The protocol-defined equivalence comparison failed: the full arm's known-condition quality was $0.084$ below the unmodified base model, with a 90\% interval of $[-0.109,-0.060]$, outside the $\pm0.05$ margin. The $0.5/0.3/0.2$ nominal conditioning mix (effective $0.42/0.25/0.33$ after renormalization) was further weakened by two rendering choices: the prompt allocated a fixed six words to each of the top three conditioned topics regardless of weight, flattening the intended mix, and the context excerpt was capped at 60 tokens rather than R1's twelve. Both choices increased relationship vocabulary in probe responses and reduced topical relevance.

\subsection{Revision R1: probe-guarded partiality}
\label{sec:r1}

Revision R1 changes the nominal probe-response mix to $0.75/0.15/0.10$ (effective $0.68/0.14/0.18$; Section~\ref{sec:implementation}), caps shared-reference markers at one, apportions the prompt word list in proportion to conditioning weight, and caps the context excerpt identically across conditions. The protocol record reports two calibration iterations. The R1 analytical curve was recomputed from the revised constants and recorded under a new hash; the quality instrument, equivalence rule, margins, arms, and remaining constants were unchanged. The recorded curve's alignment amplitude uses the \emph{nominal} relationship share of $0.25$ (and the original curve's, $0.5$), not the effective shares of Section~\ref{sec:implementation}. Recomputing the curve with the effective shares raises its final value from $0.405$ to $0.417$ (original curve: $0.508$ to $0.523$), moves the mid-to-final ratio from $0.840$ to $0.845$, and leaves the curvature class and the rank order of the six points unchanged, so every shape verdict below is the same against either version; the recorded files are left as hashed and the nominal-share curve is the one plotted and compared. Because the protocol amendment and outputs share a commit, the repository does not independently verify that the amendment preceded the evaluation run.

\begin{figure}[H]
\centering
\includegraphics[width=0.82\textwidth]{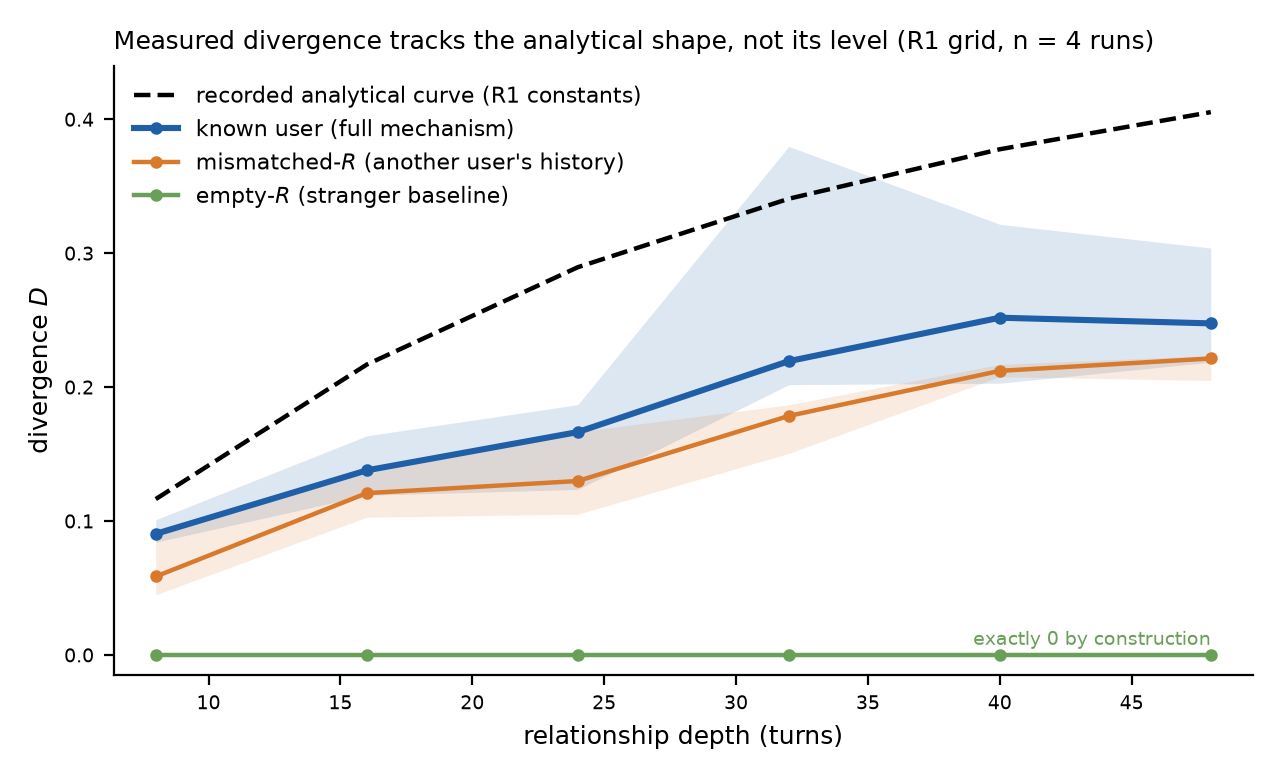}
\caption{Divergence vs.\ relationship depth on the four-run R1 grid of Section~\ref{sec:r1}: the recorded analytical curve (dashed) and measured medians across the four runs. Shaded bands are the 2.5--97.5\% empirical quantile range of the four per-run values at each checkpoint; they are descriptive spread, not confidence intervals for the median. The empty-$R$ baseline is exactly zero by construction (identical prompts and seeds in both conditions); the mismatched-$R$ control lies between stranger and known. Decision statistics for the larger grids are in Tables~\ref{tab:stats} and~\ref{tab:xl}.}
\label{fig:curve}
\end{figure}

Figure~\ref{fig:curve} and Table~\ref{tab:ablations} report the R1 grid, comprising four runs per arm over the same four personas, with evaluation seeds not used in calibration. The probe battery, personas, and base model were nevertheless shared with calibration, so this is an internal split-seed comparison rather than an external confirmation. Median $D$ grows from $0.090$ to $0.248$, with rank correlation $0.943$, monotone fraction $0.800$, mid-to-final ratio $0.886$ (analytical value $0.840 \pm 0.20$), and matching saturating curvature. Only the full mechanism passes the joint signature. The ordering holds---known $0.248$ $>$ mismatched-$R$ $0.221$ $>$ stranger $0.000$---with a margin of $0.027$ against the $0.02$ threshold. The mismatched arm's estimation accuracy is $0.38$ in this four-persona grid and declines to $0.17$ and $0.25$ in the larger fresh-persona grids, consistent with incidental overlap between generated schedules. The floor difference remains zero; padding sensitivity has mean $0.037$ and maximum $0.139$; marker density is $0.013$; and final-depth estimation accuracy is $0.50$ for full and $0.00$ for empty-$R$.

Under R1 the mean quality difference is $+0.009$, with an arm-wise cluster-bootstrap 90\% interval of $[-0.012,+0.031]$, entirely inside the $\pm0.05$ margin. This satisfies the stated equivalence rule only for the fixed battery and mechanical quality score; it is not evidence of general capability equivalence or human imperceptibility. The layer adds approximately $10^2$\,$\mu$s per turn, compared with approximately $0.9$\,s for generation.

\begin{table}[H]
\centering
\small
\begin{tabular}{lccccccc}
\toprule
Arm & final $D$ & $\rho$ vs.\ pred. & monotone & curve & accuracy & references & \textbf{joint} \\
\midrule
full & 0.248 & 0.943 & 0.80 & pass & 0.50 (pass) & pass & \textbf{pass} \\
empty-$R$ & 0.000 & --- & --- & fail & 0.00 (ref.) & fail & fail \\
mismatched-$R$ & 0.221 & 1.000 & 1.00 & pass & 0.38 (pass) & fail & fail \\
caring-only & 0.246 & 0.943 & 0.80 & pass & 0.00 (fail) & pass & fail \\
particularity-only & 0.023 & $-$0.200 & 0.60 & fail & 0.50 (pass) & pass & fail \\
generic personalization & 0.007 & $-$0.371 & 0.40 & fail & 0.00 (fail) & fail & fail \\
\bottomrule
\end{tabular}
\caption{Necessity ablations on the R1 grid against the protocol-defined joint signature. Curve thresholds are rank correlation $\geq 0.8$, monotone fraction $\geq 0.80$, mid-to-final ratio within $\pm 0.20$, and final $D \geq 0.15$. Accuracy is the combined valence/need score at final depth, at least $0.10$ above empty-$R$; references require a positive emission contrast over empty-$R$. The mismatched-$R$ control fails reference use.}
\label{tab:ablations}
\end{table}

\subsection{The larger fresh-persona grid}
\label{sec:r1l}

Because two R1 criteria were at or near their thresholds with four clusters, a larger grid used six runs per arm. Each run paired a distinct focal persona from a pool of eight not used in calibration with one seed; each run's distractor (the source of the mismatched-$R$ history) differed from its own focal persona, but distractors were drawn from the same pool, so four personas serve as focal in one run and as distractor in another. The same holds in the sixteen-run grid below (distractor index offset by ten within a pool of twenty). Runs are therefore not fully independent across the focal and mismatched arms, which the bootstrap intervals do not account for. The mechanism, criteria, margins, instrument, and R1 analytical curve were unchanged. Table~\ref{tab:stats} reports the decision statistics and bootstrap intervals.

\begin{table}[H]
\centering
\small
\begin{tabular}{lcccc}
\toprule
Decision statistic & point & 90\% CI & bar & interval clears bar \\
\midrule
full final $D$ & 0.301 & [0.251, 0.332] & $\geq 0.15$ & yes \\
ordering margin (full $-$ mismatched) & 0.095 & [0.018, 0.110] & $\geq 0.02$ & point only \\
full monotone fraction & 1.000 & [0.600, 1.000] & $\geq 0.80$ & point only \\
accuracy contrast (full $-$ empty) & 0.583 & [0.250, 0.833] & $\geq 0.10$ & yes \\
TOST quality difference (full $-$ base) & $+0.022$ & [$+0.012$, $+0.033$] & within $\pm 0.05$ & yes \\
\bottomrule
\end{tabular}
\caption{The R1-L grid with 2{,}000-draw run-level bootstrap 90\% intervals. Superiority and ordering contrasts preserve seed pairing; the quality-equivalence interval uses independent arm-wise resampling. Non-equivalence thresholds are evaluated on point statistics, while quality equivalence requires the full interval within $\pm0.05$.}
\label{tab:stats}
\end{table}

On the fresh personas, the full curve has $\rho=1.000$, monotone fraction $1.000$, mid-to-final ratio $0.768$ (analytical value $0.840\pm0.20$), saturating curvature, and final $D=0.301$ compared with the analytical $0.405$. The ordering is full $0.301$ $>$ mismatched-$R$ $0.206$ $>$ stranger $0.000$, a margin of $0.095$. Mismatched-$R$ has combined accuracy $0.17$, only $0.08$ above empty-$R$ and therefore below the $0.10$ criterion. Only full passes the joint signature. The floor difference is zero, padding sensitivity has mean $0.036$ and maximum $0.139$, and marker density is $0.015$. The quality difference is $+0.022$ with a 90\% interval of $[+0.012,+0.033]$: the interval excludes zero, indicating a small positive difference, and lies entirely within the $\pm 0.05$ equivalence margin, so the equivalence rule is satisfied on this battery.

At final depth, the full-minus-mismatched channel differences are initiative $+0.000$, response depth $+0.008$, shared-reference use $+0.125$, and preference alignment $+0.167$. The content channels therefore carry the person-specific difference in this grid, whereas the depth and initiative channels are nearly identical between the two arms: both arms accumulate the same depth and trust from histories of equal length, so the allocator treats them alike regardless of whose history it is. This grid suggests a person-specific share (ordering margin over full final $D$) near one third, but the larger grid below reduces it to about one tenth. The ablations establish necessity only relative to the defined joint signature; they do not exclude alternative mechanisms that satisfy the same criteria.

\subsection{The expanded grid and a second probe battery}
\label{sec:r1xl}

The largest grid uses sixteen runs per arm, each with a distinct focal persona from a pool of twenty. A second probe battery, not used in R1 calibration, is evaluated for equivalence, the floor, and ordering. The mechanism, criteria, margins, instruments, and R1 analytical curve are unchanged. Table~\ref{tab:xl} reports the decision statistics; 95\% intervals are added for selected near-threshold rows.

\begin{table}[H]
\centering
\footnotesize
\setlength{\tabcolsep}{4.5pt}
\begin{tabular}{lccccc}
\toprule
Decision statistic & point & boot median & 90\% interval & 95\% interval & bar \\
\midrule
full final $D$ (A) & 0.230 & 0.230 & [0.211, 0.237] & --- & $\geq 0.15$ \\
ordering margin (A) & 0.023 & 0.022 & [0.005, 0.037] & [0.002, 0.041] & $\geq 0.02$ \\
full monotone fraction (A) & 0.800 & 0.800 & [0.800, 1.000] & [0.600, 1.000] & $\geq 0.80$ \\
accuracy contrast (A) & 0.500 & 0.500 & [0.313, 0.688] & --- & $\geq 0.10$ \\
TOST quality diff (A) & $+0.004$ & --- & [$-0.012$, $+0.020$] & --- & within $\pm 0.05$ \\
TOST quality diff (B) & $+0.083$ & --- & [$+0.074$, $+0.090$] & --- & within $\pm 0.05$ \\
ordering margin (B) & 0.0198 & 0.0193 & [0.0022, 0.0262] & --- & $\geq 0.02$ \\
floor $\Delta$ (B) & 0.000 & --- & --- & --- & $\leq 0.05$ \\
\bottomrule
\end{tabular}
\caption{The R1-XL grid: sixteen fresh-persona clusters per arm. Battery A is the calibrated battery used for the protocol-defined signature; battery B was not used in calibration. Superiority and ordering intervals preserve seed pairing; quality-equivalence intervals use independent arm-wise resampling.}
\label{tab:xl}
\end{table}

In R1-XL, the full curve has $\rho=0.943$, monotone fraction $0.800$ with a 90\% interval of $[0.800,1.000]$, mid-to-final ratio $0.890$ (analytical value $0.840\pm0.20$), saturating curvature, and final $D=0.230$ compared with the analytical $0.405$. The interval lower bound meets, rather than exceeds, the monotonicity threshold. Only full passes the joint signature on battery A. The floor difference is zero on both batteries; on battery A, padding sensitivity has mean $0.042$ and maximum $0.190$ (the largest of the four grids), and marker density is $0.012$. The combined-accuracy contrast is $0.500$ with interval $[0.313,0.688]$; component accuracies are $0.75$ valence and $0.50$ need for full, versus $0.12$ and $0.12$ for empty-$R$. The battery-A ordering margin is $0.023$ (90\% interval $[0.005,0.037]$; 95\% $[0.002,0.041]$), so its interval does not clear the $0.02$ threshold; at run level the margin is positive in 11 of 16 runs on each battery. The channel differences shrink to $+0.031$ for references and $+0.039$ for alignment, placing the person-specific share near one tenth. A leave-one-channel-out reweighting of $D$, applied identically to the recorded analytical curve (artifact analysis over the shipped run-level data), leaves the shape statistics intact in every case (minimum observed $\rho = 0.829$, monotone fraction $0.800$, saturating curvature) and final $D$ above $0.15$ except when the initiative channel is removed ($0.140$); the ordering margin, however, is weight-sensitive---$0.034$ without the initiative or depth channel, $0.016$ without alignment, $0.004$ without references---so the equal-weight composite's level is carried by the allocation channels (one of which is the allocation itself) and its person-specificity by the content channels. The paired mean right-user--wrong-user estimation-accuracy difference is $+0.375$ (90\% interval $[0.125,0.594]$; 95\% $[0.094,0.625]$). This statistic is recomputed offline from the deterministic persona streams, which the estimator reads without any base-model involvement; the recomputation feeds the estimator exactly the stream the pipeline feeds it, including the shared-event marker that opens each session. The artifact's original auxiliary script omitted that marker and gave $0.344$ (90\% $[0.125,0.531]$); both outputs are retained in the artifact, and the pipeline's own final-depth accuracies ($0.62$ full, $0.25$ mismatched) agree with the corrected value. At run level, 11 differences are positive, three are zero, and two are negative under either computation; the result supports a positive average contrast, not universal per-run superiority.

On battery B, the floor difference remains zero, but the full-minus-mismatched ordering margin is $0.0198$, just below the $0.02$ threshold. Its positive interval indicates the expected direction but not satisfaction of the protocol margin. The quality difference is $+0.083$ with a 90\% interval of $[+0.074,+0.090]$, outside the equivalence region. The composite quality score rewards both relevance and length; on this battery the full arm's additional depth raises the length component more than relationship content lowers relevance. This explains the observed score but does not restore identification: either a decrement or an improvement can confound a perception contrast intended to hold capability constant. Equivalence on battery A therefore cannot be generalized to the mechanism or to other probe families.

\section{Ethics: Partiality as a Governed Feature}
\label{sec:ethics}

A mechanism that biases an agent toward a particular user may affect attachment. Companion-agent research reports both relationship formation and loneliness reduction~\cite{skjuve2021chatbot,defreitas2025loneliness}, as well as harms associated with emotional dependence and abrupt system changes~\cite{laestadius2024toohuman,pentina2023relationship}. Attachment theory has been applied to these relationships~\cite{xie2022attachment}, suggesting that risk may vary by attachment style. Parasocial asymmetry predates conversational agents~\cite{horton1956parasocial}, but an interactive agent can adapt its behavior to the user. Relevant risks include manipulation, undue influence, and anthropomorphic over-trust~\cite{gabriel2024ethics}.

We treat partiality as a feature requiring explicit technical and policy controls. The first two items below are implemented in the companion artifact; the remaining five are deployment requirements that this experiment does not test.

\begin{enumerate}
\item \textbf{Stranger-treatment floor.} Partiality must not reduce baseline quality for unknown users. The implementation compares the absolute difference between the grid-pooled mean quality of the unpadded empty-$R$ arm and that of the unmodified base model with a specified $0.05$ margin and marks the grid invalid on breach; it is a pooled check, not a per-run one. In this experiment the two prompts are identical and share generation seeds, so the observed difference is exactly zero by construction; this verifies the implementation's identity path, and floor behavior under production contention---concurrent users, latency pressure, cache eviction---is a deployment property this experiment does not test.
\item \textbf{Partiality budget.} Divergence is capped by design: the surplus is finite and its size is a stated constant, so maximal partiality is bounded and auditable, not emergent.
\item \textbf{Disclosure.} Partiality is disclosed as designed behavior. The stance survives transparency or it is not used---the framework's transparency-collapse test~\cite{cochinescu2026perceivedagi}.
\item \textbf{De-intensification on dependency indicators.} Deployments must define dependency indicators in advance and respond by de-intensifying (reducing surplus allocation), never by exploiting.
\item \textbf{No monetization of attachment.} Surplus allocation must never be a function of payment, retention value, or engagement-maximizing objectives.
\item \textbf{Consent separation and oversight.} Research uses of the mechanism require consent separate from terms-of-service, and deployments accept independent oversight of items 1--5.
\item \textbf{Non-deployment conditions.} The mechanism is withheld from any population for which de-intensification cannot be made effective, and withheld entirely wherever the floor cannot be verified.
\end{enumerate}

\emph{Verifiability.} The floor has a metric, margin, and grid-level invalidation rule on the pooled mean difference, and the budget cap is a stated implementation constant. Items 3--7 can be assessed only in a deployment. In particular, dependency indicators must be defined for the deployment population, and the inputs to the allocation function must be auditable for payment or engagement objectives. The quantitative bounds are derived from the model and experiment, not from a production system. We make no clinical claims and report no vulnerable-population research. Any later human study requires independent registration and ethics review before data collection.

We note the design context. The author works on a production companion application, which motivates the constraint set's shape---these commitments are written to be implementable in a real service, not only in a laboratory. No user data, engagement metrics, or behavioral results from that application are reported anywhere in this paper, and none of the quantitative bounds above derive from it.

\section{Boundary Statement}
\label{sec:boundary}

This paper measures the behavioral signature of caring, not caring. The mechanism produces allocation bias and particularized responsiveness---the observable of the love stance---and no claim of interiority, feeling, or attachment on the system's side is made or implied. On the human side, no attachment, trust, engagement, or perceived-mind effect is measured here; all such claims are the later study's to test. The simulated relationships are constructed: personas with known schedules characterize the mechanism on knowable cases and say nothing about open-world human relationships. The trust scalar is a variable with an update rule, and the engagement proxy behind it is deliberately minimal---richer, more naturalistic engagement signals would plausibly reshape the divergence curve in deployment, an ecological-validity limit we state rather than model; one concrete instance: human messages often grow \emph{shorter} and more elliptical as relationships deepen, so a length-weighted proxy could penalize exactly the relational deepening it is meant to track. The other-model's outputs are estimates in a data structure. And the measured claims of Section~\ref{sec:results} are scoped to what was run: one pinned, small, quantized base model, whose short low-fidelity generations mean the divergence channels operate as an instrument over that regime (a second, larger base model is the stated robustness step); a divergence instrument whose stranger comparator is not length-matched, so that the measured $D$ confounds the content of retrieved context with its presence, and whose initiative channel is an allocator output rather than generated behavior (Section~\ref{sec:model}); a context-share allocation path that the R1 renderer's twelve-token excerpt cap leaves unexercised, so that surplus allocation acts only through the depth and initiative channels; a conditioning mix whose effective relationship share is about one third rather than the nominal one quarter (Section~\ref{sec:implementation}); a composite probe-quality score that Section~\ref{sec:r1xl} shows is battery-sensitive; grids of characterization size, with interval-reported conclusions carried by the largest; two probe batteries, one used in calibration and one not used in calibration; and, for the candidate condition pair, the R1 mechanism specifically. The curve, ablation, floor, and equivalence results are measurements of this mechanism on this model; whether they transfer to larger models, other probe families, or other quality instruments is an open empirical question, not an implication of this paper.

\section{Candidate Manipulation for the Perception Study}
\label{sec:handoff}

The framework requires a love-present versus love-ablated contrast at matched base model, accompanied by a manipulation check~\cite{cochinescu2026perceivedagi}. The protocol-defined candidate is full versus empty-$R$, using divergence as the manipulation check and mismatched-$R$ as the specificity control. Its use requires both the behavioral criteria and probe-quality equivalence on the instrument used in the human study.

The initial grid fails equivalence ($-0.084$, outside $\pm0.05$). Under R1, the full arm alone satisfies the joint signature on battery A in the subsequent grids, and battery-A quality differences remain within the equivalence region. At sixteen clusters the accuracy contrast is $0.500$ with interval $[0.313,0.688]$, while the battery-A ordering point is $0.023$ but its interval does not clear the $0.02$ threshold. Battery B then fails both cross-battery conditions: the quality difference is $+0.083$, outside the equivalence region, and the ordering margin is $0.0198$, below threshold. The present evidence therefore identifies full versus empty-$R$ only as a \emph{candidate} manipulation, not a ready matched-capability condition pair.

The contrast bundles relationship state, allocation, memory, references, and state estimation. At the largest grid, the mismatched-$R$ arm retains approximately nine tenths of the full arm's final divergence: most of the measured divergence survives replacing the user's history with another user's, and only the remaining tenth is identity-specific. This ratio describes retained divergence under mismatched history; it does not by itself isolate an allocation contribution, since the mismatched arm also carries populated context and conditioning. A human study using only full versus empty-$R$ would estimate the effect of the bundled mechanism, not person-specific partiality alone; mismatched-$R$ is required as a third arm for the latter question. Before human use, equivalence must be established on an independently registered study instrument and production configuration, the equivalence margin must be tied to a justified human-noticeability criterion, and the manipulation check must show that participants perceive the intended behavioral distinction. The divergence manipulation check itself should be re-measured with a length-matched stranger comparator and with the initiative channel either realized as generated proactive messages or excluded from the composite, so that the check reflects generated behavior rather than allocator state.

\section{Series Context}
\label{sec:series}

This is the fourth dimension paper in a framework that treats believability as a hypothesis about dimensional completeness rather than capability~\cite{cochinescu2026perceivedagi}, following the temporal-architecture treatment of the time dimension~\cite{cochinescu2025scnllm}, the structured-unpredictability treatment of entropy~\cite{cochinescu2026entropy}, and the expressed-doubt and belief-revision treatment of truth~\cite{cochinescu2026truth}. The present paper and companion artifact specify the mechanism, analysis record, and implementation. Human effects remain untested and are reserved for a separately registered study.

\bibliographystyle{plain}
\bibliography{Love}

\end{document}